\documentclass[runningheads,fleqn]{svmult}
\usepackage{makeidx}   
\usepackage{graphicx}  
\usepackage{subeqnar}  
\usepackage{multicol}  
\usepackage{taphys}    
\makeindex             
\newcommand{\greeksym}[1]{{\usefont{U}{psy}{m}{n}#1}}
\newcommand{\umu}{\mbox{\greeksym{m}}}

\begin{document}
\title*{Intrinsic Spin Hall Effect}
\toctitle{Intrinsic Spin Hall Effect}
\titlerunning{Intrinsic Spin Hall Effect}
\author{Shuichi Murakami\inst{1}}
\authorrunning{Shuichi Murakami}
\institute{Department of Applied Physics, University of Tokyo,
Hongo, Bunkyo-ku, Tokyo 113-8656, Japan}

\maketitle              

\begin{abstract}
A brief review is given on a spin Hall effect, where an external 
electric field induces a transverse spin current.
It has been recognized over 30 years that such 
effect occurs due to impurities in the presence of spin-orbit 
coupling. Meanwhile, it was proposed recently that there is also an intrinsic 
contribution for this effect. We explain the mechanism for this intrinsic 
spin Hall effect. We also discuss recent experimental observations of
the spin Hall effect.
\end{abstract}


\section{Introduction}
In the emerging field of spintronics \cite{wolf2001}, 
it is important to 
understand the nature of spins and spin current inside 
semiconductors. There have been many proposals for semiconductor 
spintronics devices, whereas their realization remains elusive.
One of the largest obstacles is an efficient spin injection into 
semiconductors. One way is to make semiconductors ferromagnetic,
such as (Ga, Mn)As \cite{ohno1998}. 
The Curie temperature is, however, still lower than
the room temperature, and there are still rooms for improvement towards
practical use.

Spin Hall effect (SHE) can be an alternative way for efficiently
injecting spin current into semiconductors. 
In the first proposal of the spin Hall effect 
by D'yakonov and Perel \cite{dyakonov1971}, 
followed by several papers \cite{hirsch1999,sfzhang2000},
the SHE has been considered as an extrinsic effect,
due to impurities in the presence of spin-orbit (SO) coupling.
Nevertheless, quantitative estimate for this extrinsic
SHE is difficult, and this extrinsic effect is not easily controllable.

In 2003 two groups independently proposed an intrinsic spin Hall effect
in different systems. Murakami, Nagaosa, and Zhang \cite{murakami2003}
proposed it 
in p-type semiconductors like p-GaAs. 
On the other hand, Sinova et al.\ \cite{sinova2004} 
proposed a spin Hall effect in n-type semiconductors in 
two-dimensional heterostructures. 
This induced spin current 
is dissipationless, and can flow even in nonmagnetic materials. 
It shares some features in common with the quantum Hall effect.

Because the predicted effect is large enough to be
measured, even at room temperature in principle, 
this intrinsic SHE attracted much attention, and many related works,
have been done. Nevertheless there remain several
issues, relevant also for experiments.
One of the important questions is disorder effect. 
While there are a lot of works on disorder effect, the most 
striking one is by Inoue et al.\ 
\cite{inoue2003,inoue2004}.
They considered dilutely distributed impurities with short-ranged 
potentials, and calculate the SHE, incorporating 
the vertex correction in the ladder approximation. Remarkably the
resulting spin Hall conductivity is exactly zero in the clean limit.
This work made many people to consider that the SHE
is ``fragile'' to impurities; namely, only a small amount of
impurities will completely kill the intrinsic 
SHE. However, this is not in general 
true. In fact, the spin Hall conductivity  
is in general nonzero even in the presence of disorder, as we see later.

In such circumstances, two seminal experiments on 
the SHE have been done. Kato et al.\ \cite{kato2004}
observed spin accumulation in n-type GaAs by means of Kerr rotation.
Wunderlich et al.\ \cite{wunderlich2005} observed a circularly
polarized light emitted from a light-emitting diode (LED) structure,
confirming the SHE in p-type semiconductors.
Separation between the intrinsic and extrinsic SHE for these experimental
data is not straightforward, and is still under debate.

The paper is organized as follows. In Sect.\ 2 we explain 
basic mechanisms and features for the intrinsic SHE. Section 3
is devoted to a disorder effect on the SHE. In Section 4 we 
collect a number of recent interesting topics on the SHE.
In Section 5 we introduce two recent experimental reports on 
the SHE. We conclude the paper in Sect.\ 6.

\section{Intrinsic spin Hall effect}
In this section we explain the two theoretical proposals 
for the intrinsic SHE.
\subsection{Spin Hall effect in p-type semiconductors}
We begin with semiclassical description of 
the SHE, and apply it to the p-type semiconductors \cite{murakami2003}.
In this description, we introduce a ``Berry phase in 
momentum space''.  The Berry phase \cite{berry1984,shaperewilczek,niu}
is a change of a phase of a quantum state
caused by an adiabatic change of some parameters. 
As we explain later, 
Berry phase in momentum space gives rise to the Hall effect,
as first demonstrated for the quantum Hall effect
\cite{thouless1982,kohmoto1985,sundaram1999}. Here, 
the wavevector $\vec{k}$ is regarded as
adiabatically changing due to a small external electric field.
In two-dimensional systems, for example, the Hall conductivity 
$\sigma_{xy}$ in a clean system 
is calculated from the Kubo formula as 
\begin{equation}
\sigma_{xy} = -\frac{e^{2}}{2\pi h}\sum_{n}\int_{{\rm BZ}}d^{2}k\ n_{F}(
\epsilon_{n}(\vec{k})) B_{nz}(\vec{k}),
\label{sigmaxy}
\end{equation}
where $n$ is the band index, and the integral is over the entire Brillouin
zone.
$B_{nz}(\vec{k})$ is defined as a $z$ component of 
$\vec{B}_{ n }(\vec{k}) =\nabla_{\vec{k}}\times \vec{A}_{n}(\vec{k})$, where
\begin{equation}
A_{n i}(\vec{k})=-i\left\langle n\vec{k}\left|\frac{\partial}{\partial
k_{i}}\right|n\vec{k}\right\rangle\equiv -i\int_{{\rm unit\ cell}}
 u_{n\vec{k}}^{\dagger}
\frac{\partial u_{n\vec{k}}}{\partial k_{i}}d^{2}x,
\end{equation}
and $u_{n\vec{k}}(\vec{x})$ is the periodic part of the Bloch wavefunction
$\phi_{n\vec{k}}(\vec{x})=
e^{i\vec{k}\cdot\vec{x}}u_{n\vec{k}}(\vec{x})$. This $B_{nz}(\vec{k})$ 
represents the effect of Berry phase in momentum space.
$n_{F}(\epsilon_{n}(\vec{k}))$ is the Fermi distribution function for the
$n$-th band.
This intrinsic Hall conductivity (\ref{sigmaxy})
was first recognized in 
the paper by Karplus and Luttinger \cite{karplus1954}.
This Berry phase in momentum space has been studied in 
the recent works on anomalous Hall effect (AHE)
\cite{ye1999,ohgushi2000,%
taguchi2001,jungwirth2002,fang2003,yao2004}, as well 
as those on the SHE.

By incorporating the effect of $\vec{B}(\vec{k})$,
the Boltzmann-type semiclassical equation of motion (SEOM) 
acquires an
additional term  \cite{sundaram1999}:
\begin{equation}
\dot{\vec{x}}=\frac{1}{\hbar}
\frac{\partial E_{n}(\vec{k})}{\partial \vec{k}}+
{\dot{\vec{k} }}\times {\vec{B}}_{n}
(\vec{k}),
\ \ \ \ 
\hbar\dot{\vec{k}}=-e(\vec{E}+\dot{\vec{x}}\times \vec{B}(\vec{x})).
 \label{dotk}
\end{equation}
The term $\dot{\vec{k}}\times \vec{B}_{n}(\vec{x})$ represents the effect of 
Berry phase, and it is called an anomalous velocity. Under the external
electric field, the anomalous velocity becomes perpendicular to the 
field, and gives rise to the Hall effect. 
This Hall current is
distinct from the usual Ohmic current, which comes from the 
shift of the Fermi surface from its equilibrium.
This Hall effect comes from all the occupied states, not only 
from the states on the Fermi level. By summing up the anomalous velocity 
over the filled states, 
one can reproduce the Kubo formula result (\ref{sigmaxy}). 
Given the Hamiltonian, the vector field $\vec{B}(\vec{k})$ is calculable, 
and we can get the intrinsic Hall conductivity, 
as in the ab initio calculation of the AHE in \cite{fang2003,yao2004}.
Due to remarkable similarity of the two equations in Eq.~(\ref{dotk}),
$\vec{B}_{n}(\vec{k})$ can 
be regarded as a ``magnetic field in $\vec{k}$-space.
$\vec{B}_{n}(\vec{k})$ can have monopoles, and such monopoles can give nontrivial 
topological structure for magnetic superconductors \cite{murakami2003b}.

\begin{figure}[h]
\includegraphics[scale=0.77]{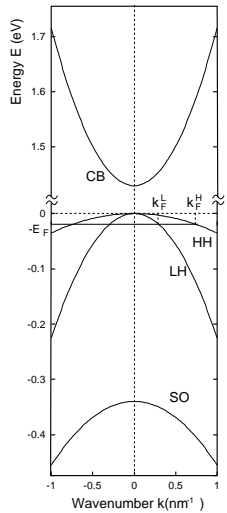}
\caption{Schematic band structure for GaAs. CB, HH, LH, SO represent  %
the conduction, heavy-hole, light-hole and split-off bands, respectively.} %
\end{figure}
This anomalous velocity leads to the SHE
in semiconductors with diamond structure (e.g. Si) or zincblende
structure (e.g. GaAs). The valence bands consist 
of two doubly degenerate bands: heavy-hole (HH) and light-hole (LH) bands.
They are degenerate at $\vec{k}=0$ as shown in Fig.~1.
Near $\vec{k}=0$, the valence bands are described by 
the Luttinger Hamiltonian \cite{luttinger1956}
\begin{equation}
H=\frac{\hbar^{2}}{2m}\left[\left(\gamma_{1}+\frac{5}{2}\gamma_{2}\right)
k^{2}-2\gamma_{2}(\vec{k}\cdot\vec{S})^{2}\right],
\end{equation}
where $\vec{S}$ is the spin-$3/2$ matrices representing the total 
angular momentum. 
For simplicity, we 
employed the spherical approximation for the Luttinger Hamiltonian,
while a calculation without it is also possible \cite{bernevig2003}.
In this Hamiltonian, a helicity $\lambda=\frac{\vec{k}\cdot
\vec{S}}{k}$ is a good quantum number, and can be used as a label
for eigenstates. The HH and LH bands have
$\lambda_{H}=\pm\frac{3}{2}$ and $\lambda_{L}=
\pm\frac{1}{2}$, respectively.
The SEOM reads as
\begin{equation}
\dot{\vec{x}}=\frac{1}{\hbar}
\frac{\partial E_{\lambda}(\vec{k})}{\partial \vec{k}}+
{\dot{\vec{k}}}\times {\vec{B}}_{\lambda}
(\vec{k}),\ \ \ 
 \hbar\dot{\vec{k}}=e\vec{E}.
\label{eq:som}
\end{equation}
Because we are considering holes, the sign of the charge has been changed.
By straightforward calculation, we get
$\vec{B}_{\lambda}
(\vec{k})=\lambda(2\lambda^{2}-\frac{7}{2})\vec{k}/k^{3}$.
Hence, the anomalous velocity due to Berry phase is along
the direction $\vec{E}\times\vec{k}$. By integration in terms of 
the time $t$, we get a trajectory of the holes as shown in Fig.~2.
This shows the motion projected on a plane perpendicular 
to $\vec{E}$.
We note that a semiclassical trajectory can be calculated 
directly from the Heisenberg equation of motion; the resulting trajectory
agree well with the one from (\ref{eq:som}), but with small rapid oscillations 
\cite{jiang2004}. This justifies validity of 
the SEOM (\ref{eq:som}) for adiabatic transport, which 
was questioned in \cite{wang2005}.
\begin{figure}[h]
\includegraphics[scale=0.5]{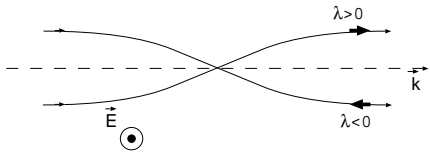}
\caption{Trajectory of holes from the semiclassical equation %
of motion with Berry-phase terms. %
This is a projection on the plane perpendicular to the electric field %
$\vec{E}$. The transverse shift of the trajectory is to the opposite %
direction, depending on the sign of the helicity %
$\lambda=\hat{k}\cdot\vec{S}$. The bold arrows represent %
the direction of spin $\vec{S}$}
\end{figure}

Due to the anomalous velocity, the motion of the holes is deflected from 
an otherwise straight motion along $\vec{k}$ (dashed line). 
The shift of the motion is opposite for the signs of the helicity $\lambda$,
referring to whether the spin $\vec{S}$ and the wavevector $\vec{k}$ are 
parallel or antiparallel.
This shift amounts to the SHE. 
By summing up this shift over the 
occupied states, 
we can calculate a spin current in response to the electric field.
If the electric field is along $l$-axis, the spin current with $S^{i}$ spin 
flowing toward the $j$-direction is \cite{murakami2003}
\begin{equation}
j_{j}^{i}=\frac{e}{12\pi^{2}}(3k_{F}^{{\rm H}}-k_{F}^{{\rm L}})\epsilon_{ijl}
E_{l},
\label{spincurrent}
\end{equation}
where $k_{F}^{{\rm H}}$ and $k_{F}^{{\rm L}}$ are the Fermi wavenumber for 
the HH and LH bands, respectively. This is schematically shown in Fig.~3.
Nominal values obtained spin Hall conductivity for p-GaAs are
of the similar order of magnitude as the condutivity at room 
temperature \cite{murakami2003}. 
In GaAs, the nominal energy difference of the two bands is larger than  
the room temperature, and the effect can in principle survive even
at room temperature.
\begin{figure}[h]
\includegraphics[scale=0.5]{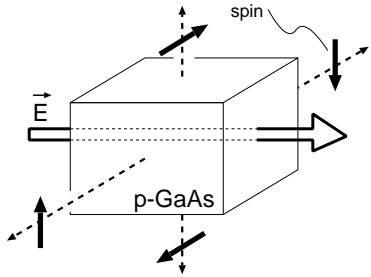}
\caption{Schematic of the spin current induced by an electric field}
\end{figure}

\subsection{Spin Hall effect in n-type semiconductors in heterostructure}
In n-type semiconductors with diamond or zincblende structure, 
the SO coupling is small. On the other hand,
however, if they are incorporated into two-dimensional heterostructure,
the inversion symmetry is broken, and the SO coupling becomes
relevant. The Hamiltonian is approximated as
\begin{equation}
H=\frac{k^{2}}{2m}+\lambda(\vec{\sigma}\times \vec{k})_{z},
\label{eq:Rashba}
\end{equation}
where $\sigma_{i}$ is the Pauli matrix.
The second term is called the Rashba term \cite{rashba1960,bychkov1984},
representing the SO coupling.
The coupling constant $\lambda$ can be experimentally determined,
and can be controlled by the gate voltage \cite{nitta1997}.

Sinova et al.\ applied the Kubo formula to this Rashba Hamiltonian 
\cite{sinova2004}. For this procedure, they defined the spin current
$J_{y}^{z}$ to be a symmetrized product of the spin $S^{z}$ and 
the velocity $v_{y}=\frac{\partial H}{\partial k_{y}}$.
By assuming no disorder, the resulting spin Hall conductivity 
is $e/(8\pi)$, 
which is independent of the Rashba coupling $\lambda$.
The Rashba term in (\ref{eq:Rashba}) can be regarded as a $\vec{k}$-dependent
effective Zeeman field $\vec{B}_{\mathrm{eff}}
=\lambda(\hat{z}\times\vec{k})$.
In an equilibrium the spins are pointing either parallel or 
antiparallel to $\vec{B}_{\mathrm{eff}}$ for the lower and upper bands, 
respectively.
An external electric field $\vec{E}\| \hat{x}$ 
changes the wavevectors $\vec{k}$ of 
Bloch wavefunctions, and $\vec{B}_{\mathrm{eff}}$ also changes
accordingly.
The spins will then precess around $\vec{B}_{\mathrm{eff}}$,
and tilt to the $\pm z$-direction, depending on the sign of $k_{y}$.
This appears as the SHE, and the spin Hall conductivity is
calculated to be 
$\frac{e}{8\pi}$, in agreement with the Kubo formula.
We can incorporate also the Dresselhaus term representing the 
bulk inversion-symmetry breaking of the zincblende structure. 
The Hamiltonian becomes 
\begin{equation}
H=\frac{k^{2}}{2m}+\lambda(k_{x}\sigma_{y}-k_{y}\sigma_{x})
+\beta(k_{x}\sigma_{x}-k_{y}\sigma_{y}).
\end{equation}
The spin Hall conductivity $\sigma_{s}$ is given as follows;
$\sigma_{s}=e/(8\pi)$ for $\lambda^2>\beta^2$, and
$\sigma_{s}=-e/(8\pi)$ for $\lambda^2<\beta^2$ \cite{shen2004a,sinitsyn2004}.

When a perpendicular magnetic field is applied, the spin Hall conductance
will have a resonant behavior as a function of the magnetic field
\cite{shen2004b,shen2004c}. The Zeeman splitting induces  degeneracies between 
different Landau levels, and if this degeneracy 
occurs at the Fermi level, the spin Hall conductance is divergent.
An in-plane magnetic field also affects the SHE, as studied in 
\cite{chang2005}

Close relationship between the SHE 
and the Pauli spin susceptibility \cite{dimitrova2004b,erlingsson2005,shekhter2005} 
or the dielectric function \cite{rashba2004a} has been argued for these
models. An effect of electron-electron interaction has been investigated by use of this
relationship \cite{shekhter2005}.
In the Luttinger model, on the other hand, 
these three have different frequency dependence \cite{bernevig2005a}.
Thus it is not clear whether this relationship 
remains for disordered case or for generic SO-coupled models.

\subsection{General properties of the intrinsic spin Hall effect}
We consider the intrinsic SHE
to be robust against spin relaxation.
Momentum relaxation induces rapid spin relaxation
via the SO coupling. Namely, if the Fermi surface deviates
from its equilibrium position, the momentum and spin distributions
will rapidly relax. Nevertheless, because the spin Hall current comes 
from the anomalous velocity, it will survive even when the momentum
and spin relaxation is in equilibrium.
On the other hand, near sample boundaries, the spin Hall current 
will induce spin accumulation. Hence, the spin distribution 
is deviated largely from equilibrium, and the spin relaxation 
becomes effective. The amount of accumulated spins
is roughly estimated as a product of spin current and spin relaxation 
time $\tau_s$ \cite{murakami2003}. The spin accumulation 
affects also the spin current itself near the boundaries \cite{ma2004,hu2004}.

Because spin current is even under time-reversal, it can be
induced even when the time-reversal symmetry is preserved.
It also implies that the spin Hall current is dissipationless.
In doped semiconductors, however, the longitudinal conductivity is
finite and the system undergoes Joule heating. Nevertheless, 
there exist spin-Hall insulators, which are band insulators with nonzero SHE; 
in such systems the longitudinal conductivity is zero, and the SHE accompanies
no dissipation.

\section{Disorder effect and extrinsic spin Hall effect}
One can include an effect of the self-energy broadening $1/\tau$
by disorder \cite{schliemann2004,schliemann2005}. The intrinsic SHE is
reduced as expected. 
In the clean limit the spin Hall conductivity reproduces its 
intrinsic value. 
Inoue et al.\ \cite{inoue2003,inoue2004}
found an important result. They assumed 
dilutely distributed impurities with a $\delta$-function 
potential.  They calculated the self-energy within the self-consistent 
Born approximation, and the vertex correction within the ladder 
approximation. In a clean limit, 
they obtained a vertex-correction contribution 
$-\frac{e}{8\pi}$ to the spin Hall conductivity, 
exactly cancelling the intrinsic value $\frac{e}{8\pi}$.
This result was rediscovered and generalized by several people
\cite{mishchenko2004,dimitrova2004a,chalaev2004,khaetskii2004,raimondi2005,rashba2004b,malshkov2005,liu2004,liu2005a,liu2005b,sugimoto2005}.
In particular, by the Keldysh formalism,
it is found that the spin Hall current appears {\it only near the
electrodes} whereas in the bulk of the sample the spin Hall current vanishes irrespective of the lifetime $\tau$ \cite{mishchenko2004}.
The Keldysh formalism is used for more generalized cases in the Rashba model
and related models \cite{liu2004,liu2005a,liu2005b,sugimoto2005}.

One may wonder whether the SHE vanishes in other systems,
and there remain some controversies in this respect.
It is well-established that 
the SHE vanishes in the Rashba model with $\delta$-function impurities
in the clean limit.
Meanwhile, even for the Rashba model,
it is still under debate whether it vanishes 
for finite $\tau$ \cite{mishchenko2004,dimitrova2004a,chalaev2004,khaetskii2004,raimondi2005,liu2004,sugimoto2005,nomura2005,sheng2005b}
or for finite-ranged impurities \cite{inoue2004,khaetskii2004,raimondi2005,liu2004}.
Here we note that the Rashba model is exceptional, in that 
the SHE vanishes rather accidentally, namely 
because the spin current operator $J_{y}^{z}$ is
proportional to $\dot{S}_{y}=i[H,S_y]$ \cite{shekhter2005,dimitrova2004a,chalaev2004,rashba2004b,liu2005a}.
In fact one can check that the SHE does not vanish in 
general models;
for example,
when the Rashba model is generalized to include a higher-order term in $k$ 
\cite{murakami2004b}, the spin Hall conductivity no longer vanishes. In 
addition, there are some models where the vertex correction does not cancel 
the intrinsic value \cite{liu2005b}, or even vanishes by
symmetry \cite{murakami2004b,bernevig2004c}.

In retrospect, extrinsic SHE has been considered since more than
thirty years ago \cite{dyakonov1971,hirsch1999,sfzhang2000},
as mentioned in the Introduction.
The relationship with the disorder effect on the intrinsic SHE is, however,
unclear at present.
We note that there have been a similar debate in the disorder effect on the AHE over decades.
To summarize, the studies on the disorder effect are so far restricted
to special models; general and exhaustive understanding for the disorder 
effect on the SHE is still lacking.

\section{Discussions}
In this section we discuss several topics on the SHE,
which are still currently
under intensive research.
\subsection{Definition of the spin current}
In the presence of the SO coupling, the total spin 
is not conserved. Hence there is no unique way to 
define a spin current. 
Naively we expect that the ``spin current'' $\vec{J}^{s}$ 
should satisfy the equation of continuity 
$\frac{\partial S_{i}}{\partial t}+\nabla\cdot
\vec{J}^{s}_{i}=0$; this relationship requires the conservation of 
total spin, namely,
\begin{equation}
0=\frac{\partial}{\partial t}\int S_{i}d^{d}x=
-i\left[\int S_{i}d^{d}x,H\right].
\end{equation}
In the cases relevant for the SHE, the SO coupling violates
this conservation of total spin.
In other words, due to the nonconservation of spin, Noether's theorem
is not applicable for a definition of spin current.

One can adopt the symmetrized product $\frac{1}{2}(v_i S_j+S_j v_i)$
between the velocity $\vec{v}$ and the spin $\vec{S}$ as a 
definition of the spin current as in 
\cite{sinova2004}. The result calculated by the Kubo formula 
with this definition is in general different
from that by semiclassical theory described above 
\cite{murakami2004a}.
This difference comes from noncommutatibity between 
the spin $\vec{S}$ and the velocity $\vec{v}$.
In other words, this comes from the non-uniqueness of the
definition of spin.
One can modify the semiclassical theory to give the same result as
the Kubo formula, by adding 
three contributions:
spin dipole, torque moment, and change of wavepacket spins due to
electric field \cite{culcer2003}.

An alternative way is to separate the spin $\vec{S}$ into conserved (intraband)
part $\vec{S}_{\mathrm{(c)}}$
and nonconserved (interband) part $\vec{S}_{\mathrm{(n)}} $\cite{murakami2004a}.
As $[\vec{S}_{{\rm (c)}},H]=0$, spin current can be uniquely defined 
for $\vec{S}_{{\rm (c)}}$. The resulting spin current is
different from 
(\ref{spincurrent}), and this difference is considered as
a quantum correction to (\ref{spincurrent}).
The reason why we only take $\vec{S}_{{\rm (c)}}$ is because
relaxation by impurities will rapidly smear out the non-conserved part
$\vec{S}_{{\rm (n)}}$.
Another attempt for defining conserved spin current is done by introducing 
torque dipole moment \cite{pingzhang2005}; this definition ensures the 
Onsager relation.
Because there is no unique definition for the spin current,
we have to choose one definition which matches the considered experimental 
setup to measure the spin current.

\subsection{Landauer-B{\"u}ttiker formalism}
The four-terminal Landauer-B{\"u}ttiker formalism can be used 
to study the SHE. 
In \cite{nikolic2004a,sheng2005a}, the authors used
a tight-binding Hamiltonian with SO coupling on a square lattice, 
and used the 
four-terminal Landauer-B{\"u}ttiker formalism to study mesoscopic SHE
for system size up to $\sim 100\times 100$.
In the bottom of the band the tight-binding 
model reduces to the Rashba model.
They first studied the SHE without any disorder, by changing 
the system size and the SO coupling.
The resulting spin Hall conductivity is not equal to the universal value 
of $\frac{e}{8\pi}$, and is critically dependent on the 
strength of the SO coupling. 
They also studied the dependence on on-site disorder.
Even in the presence of disorder, the SHE remains nonzero, and 
depends on the disorder strength. The Luttinger model was studied in a similar
fashion \cite{chen2005}.
We remark that because of the nonuniqueness of the definition of spin 
current, the comparison between 
the results by Landauer-B{\"u}ttiker formalism and those by Kubo formula
is not straightforward.

Non-equilibrium spin accumulation has been studied in a two-terminal 
geometry \cite{nikolic2004b}. The Keldysh non-equilium Green's function is
combined with the Landauer formalism to study numerically the 
spin accumulation.
The spins accumulate at the both edges, with their direction along $z$-axis
opposite for the two edges. Spin accumulation at the edges of
ballistic systems is also studied \cite{usaj2004}.
This accumulation at the edges is qualitatively similar to the experimental 
result \cite{kato2004}.

In \cite{hankiewicz2004}, an H-shaped structure is proposed 
for a measurement of the SHE via dc-transport porperties.
With use of the Landauer-B{\"u}ttiker formalism,
the dc voltage response is calculated with realistic parameters.
In \cite{souma2004},
the Landauer-B{\"u}ttiker 
formalism is applied to a mesoscopic ring with Rashba SO coupling;
by tuning the Rashba coupling, the spin Hall current oscillates 
due to the Aharanov-Casher phase around the ring.

\subsection{Criterion for nonzero SHE, spin Hall insulator}
The SHE is induced by the SO coupling, which is inherent in
every material. However, if the bands within the same multimplet are 
all filled or all empty, the bands do not contribute to the SHE. 
Thus, the criterion for nonzero SHE 
is that the bands within the same multiplet 
have different fillings. For example, in GaAs, the valence bands 
($J=3/2$) consist of the LH and the HH; therefore, p-doping 
brings about the difference of fillings between the LH and the HH, 
giving nonzero SHE. On the other hand, the conduction band ($J=1/2$)
is doubly degenerate (if we ignore the bulk inversion symmetry breaking 
from the zincblende structure), and n-doping does not give rise to
nonzero SHE. If we incorporate it into heterostructure, the degeneracy of
the conduction band is lifted, and n-doping induces nonzero SHE.

According to this criterion, some band insulators 
have nonzero SHE, even though the charge conductivity is zero 
\cite{murakami2004c}. Two classes of materials have been proposed
for such ``spin Hall insulators'' in \cite{murakami2004c}. 
One is zero-gap semiconductors 
such as HgTe and $\alpha$-Sn. By introducing uniaxial anisotropy the 
gap becomes finite.
The other is narrow-gap semiconductors such as PbTe.
In these semiconductors the gaps come from the SO coupling,
and the SHE is nonzero even though they are band insulators.
In \cite{kane2004}, on the other hand, a graphene sheet is proposed 
to be a quantum spin Hall insulator.

One can consider a Hall effect for the orbital angular momentum (OAM)
instead of spin \cite{zhangyang2005}. In the Rashba model, the 
resulting intrinsic Hall conductivity for the OAM is $-e/(8\pi)$, 
exactly cancelling the intrinsic SHE. Thus the intrinsic Hall 
effect for the {\it total angular momentum} vanishes.
It follows from the conservation of the total angular momentum 
$s_{z}+L_{z}$ \cite{zhangyang2005}, i.e. from 
a continuous rotation symmetry around the $z$-axis.
Nevertheless, for general systems
it is not true, and the cancellation does not take place in general.

Several first-principle calculations have been done \cite{guo2005,yao2005}.
In \cite{yao2005}, the intrinsic 
SHE is calculated for Si, GaAs, W and Au for 
various values of the Fermi level. It is found that even without 
doping, Si and GaAs show a small but finite SHE. It is due 
to a small hybridization.
In this sense, these undoped semiconductors are spin Hall insulators.
In \cite{guo2005}, on the other hand,
the intrinsic SHE is calculated for n-type Ge, GaAs
and AlAs as a function of the hole concentration and strain. They also 
calculated the Hall effect for the OAM, and showed that they do not
cancel with the spin Hall effect.

\section{Experiments}
Kato et al.\ \cite{kato2004} observed the SHE in n-type
semiconductors by measuring spin accumulation at the edges of the
sample by Kerr rotation.  The spin accumulation is uniformly
distributed along the both edges.  They evaluated from experimental
data the amount of spin accumulation and spin lifetime as a function
of an external magnetic field. The measured spin Hall resistivity is
2 M$\Omega\umu$m.  They concluded the observed SHE to be
extrinsic for the following reasons; (i) spin splitting is negligibly
small in the sample, and (ii) the effect has no dependence on crystal
orientation.

Nevertheless, Bernevig and Zhang argued that the observed SHE can be intrinsic,
coming from the Dresselhaus term representing bulk inversion-symmetry breaking
\cite{bernevig2004c}. They showed that even if the spin splitting
due to the Dresselhaus term is negligibly small, the SHE can be as large 
as the experimental data. It can also account for the absence of 
dependence on crystal orientation. Thus, the source of the observed SHE is 
still to be resolved.

The observed distribution of spin accumulation 
is clearly different from the Keldysh formalism calculation
on the Rashba model \cite{mishchenko2004}. As discussed in the previous 
section, the Rashba model may not be general enough to be useful for
comparison with experiments.

On the other hand,
Wunderlich et al.\ \cite{wunderlich2005} observed the SHE in a 
2D p-type system, using a p-n junction light-emitting diode.
They applied an electric field across the hole channel, and 
observed a circular polarization of the emitted light, whose sign is opposite
for the two edges of the channel.  The circular polarization is $\sim 1\%$
at maximum. They argued that it is near to the clean limit, and the obtained 
SHE is mostly intrinsic. More refined argument, by showing vanishing 
vertex correction, also supports this conclusion
\cite{bernevig2004b}. 

As we have seen in Sect.\ 3, disorder effect on the SHE is still 
under intensive studies. It will take some time to determine 
whether the above experimental results for the SHE is mostly intrinsic or
extrinsic (or both). To pursue this issue experimentally, it will be ideal to 
change systematically the disorder, and measure the SHE in the same
line as in \cite{lee2004}.

\section{Summary}
In this review we summarize recent results for the SHE. 
In these two years there has been much progress in this field, both
in theories and in experiments. Nevertheless, as many results have 
accumulated, we come up with new questions to be solved. In the 
present stage there are a lot of ways to approach the problem 
theoretically and experimentally, and the results from different 
methods have not yet 
satistactorily converged into a unified picture of the SHE.
In particular, disorder effect is the key issue to enable comparison 
between theories and experiments in a systematic fashion.

The author would like to thank N. Nagaosa, S. Onoda, N. Sugimoto,
and S.-C. Zhang for
collaborations. He is grateful to G.~E.~W. Bauer, D. Culcer, Z. Fang,
J. Inoue, A. H. MacDonald,
B. Nikoli{\'c}, Q. Niu, K. Nomura,
M. Onoda, S.-Q. Shen and J. Sinova for fruitful discussions.
This work has been supported by a Grant-in-Aid (No.~16740167) 
for Scientific 
Reserach from the Ministry of Education, Culture, Sports, 
Science and Technology of Japan.

%

\end{document}